\documentclass[aps,prl,reprint,
onecolumn]{revtex4}
\usepackage{graphicx}
\usepackage{bm}
\usepackage[latin1]{inputenc}
\usepackage{subfigure}

\newcommand{\ba}{\begin{array}}
\newcommand{\ea}{\end{array}}
\newcommand{\be}{\begin{equation}}
\newcommand{\ee}{\end{equation}}
\newcommand{\bea}{\begin{eqnarray}}
\newcommand{\eea}{\end{eqnarray}}

\begin{document}
\title{Solvable delay model for epidemic spreading: the case of Covid-19 in Italy}

\author{Luca Dell'Anna}
\affiliation{Dipartimento di Fisica e Astronomia "G. Galilei", Universit\`a degli Studi di Padova, via F. Marzolo 8, 
35131 Padova, Italy}
\begin{abstract}
We study a simple realistic model for describing the diffusion of an infectious disease on a population of individuals. The dynamics is governed by a single functional delay differential equation, which, in the case of a large population, can be solved exactly, even in the presence of a time-dependent infection rate. This delay model has a higher degree of accuracy than that of the so-called SIR model, commonly used in epidemiology, which, instead, is formulated in terms of ordinary differential equations. We apply this model to describe the outbreak of the new infectious disease, Covid-19, in Italy, taking into account the containment measures implemented by the government in order to mitigate the spreading of the virus and the social costs for the population.
\end{abstract}
\maketitle
\section{Introduction}
In a very few months a viral infection called Covid-19 (Coronavirus disease 19) originated in China, breaking through the borders of all the countries, 
 rapidly spread all over the globalized world. Italy is one of the hardest hit countries suffering from the very dramatic consequences of this disease. 
The outbreak of the virus, the new coronavirus which caused the infection, seems out of our control. In the absence of a therapy and a vaccine, 
social distancing measures and a strict lockdown appear to be the most effective means to contain the growth of the infection. 
We should remind that there are places in the world where often infectious diseases, also those already defeated in the so-called more developed countries, can still cause very severe consequences among the local populations. 

Even if we cannot answer the question why a virus starts spreading and which is its origin, we can still wonder how it diffuses. 
The aim of this work is, therefore, to provide a simple handy model for epidemic spreading, which could depend only on the couple of parameters which generally characterize an infectious disease: the infection rate and the infectiousness (or recovery) time. Both these quantities can be taken from the experience, therefore, we do not need further parameters to fit the data which could cause artificial predictions. We will show that the model we are presenting have the same, or even higher, predictive power than that of one of the most widely used technique in epidemiology, the SIR model \cite{anderson,keeling,sirmodel}. 
This latter model requires the presence of a 
recovery rate related to the number of recovered persons, without considering that the new cases of recovery (and fatality) come from infected cases occurred previously. 
The model we are proposing, instead, is based on the fact that the closed cases come from the infected ones after an average delay recovery time, therefore, contrary to the SIR model, formulated in terms of a set ordinary differential equations, it is described by just a functional retarded differential equation, bringing predictions more under control. 
In this work we derive the exact analytical solution of this model in the limit of a large population, 
also in the presence of a time-dependent infection rate, which is the case when containment measures are implemented in order to reduce the spreading of the infection. 
Moreover, the definition of the so-called basic reproduction number ${\cal R}_0$ (a parameter determining whether a infectious disease can spread or not) comes out naturally in our delay model. 
Actually delay models in epidemiology have been already implemented in many cases \cite{diekmann, arino, zhang, beretta, ruschel, young}. We consider the case where the infection period is constant and provide for the first time an analytical result for the spreading of the disease in the early stage of the infection. 

We finally apply this technique to give a quantitative description of the diffusion of Covid-19 in Italy, showing the current scenario based on the actual situation and what would have happened without the containment measures. Generally it is quite difficult to give a reliable forecast on the fate of the  epidemic spreading because it heavily depends on individual and social behaviors, on the effectiveness of the containment measures already implemented, or that will be taken, by the government and on the future political decisions. 
At the time being, even if the situation in Italy is improving, it seems that more efforts are needed in order to change course and rapidly stop the spreading of the disease. Further measures might be useful, like, for instance, i) running more diagnostic tests, at least, on all the doctors and medical workers who are in contact with many patients, ii) improving the food distribution to avoid the crowding in the food shops and to ensure subsistence goods also to those who need, iii) providing medical devices like surgical masks to all the population. 

As last remark, we remind that the outbreak of Covid-19 has been declared a pandemic by the World Health Organization. Many countries are already heavily overwhelmed by this infection and by the risk for the public health, therefore, in a networked world we all have to behave and operate with an improved spirit of cooperation. The bitter lesson imparted by this tough situation is that we cannot save ourselves alone.

\section{Results}
\subsection{The model}

Let us introduce the model, assuming that the full population is constant, uniform, homogeneously mixed, and counts $N$ persons who can be divided in three parts, susceptible, infected and recovered persons, whose numbers, at a given time $t$, are $S(t)$, $I(t)$ and $R(t)$, respectively.\\
Let us define $I_o$ the initial infected persons at time $t=0$, and introduce ${\cal P}(t)$, the probability of remaining infectious at later time $t$ after becoming infectious. ${\cal P}(t)$ is a monotonic decreasing function with ${\cal P}(0)=1$ and $\lim_{t\rightarrow \infty}{\cal P}(t)=0$. 
The initial number of the first infectious persons, therefore, decreases according to $I_o{\cal P}(t)$, meanwhile other susceptible persons become infected after coming in contact with those already infected, with a rate of infection $\alpha$, which counts the number of contacts per unit of time, times the probability for a infected person to transmit the infection. The probability of new infections at a given time $x$ is, therefore, proportional to the ratio $S(x)/N$
of persons who are still susceptible and the number of infected persons who are still infectious, $I(x){\cal P}(t-x)$. At a later time $t$, the total number of infections are, therefore, given by
\be
I(t)=I_o{\cal P}(t)+\frac{\alpha}{N} \int_0^t S(x) I(x) {\cal P}(t-x) dx
\label{eq1}
\ee
Equivalently, writing ${\cal P}^\prime(t)=\frac{d{\cal P}(t)}{dt}$, Eq.~(\ref{eq1}) can be written as
\be
\frac{dI(t)}{dt}=
\frac{\alpha}{N} S(t) I(t)+
I_o{\cal P}^\prime(t)+\frac{\alpha}{N} \int_0^t S(x) I(x) {\cal P}^\prime(t-x) dx
\label{eqI}
\ee
Since ${\cal P}(t)$ is a non-increasing function, ${\cal P}^\prime(t)$ is negative, therefore the last two terms of Eq.~(\ref{eqI}) reduces the increase of infection due to the first term. For that reason we can identify those terms as minus the variation of the removed cases  
\be
\frac{dR(t)}{dt}=-I_o{\cal P}^\prime(t)-\frac{\alpha}{N} \int_0^t S(x) I(x) {\cal P}^\prime(t-x) dx
\label{eqR}
\ee
It is convenient, for the benefit of future discussion, to introduce the total number of infected persons, either those who are still infected at time $t$, $I(t)$,  
and those who recovered or died, $R(t)$,
\be
F(t)=I(t)+R(t)
\ee
From Eqs.~(\ref{eqI}) and (\ref{eqR}), since $S(t)+F(t)=N$, we have that $F(t)$ fulfills the following equation 
\be
\frac{dF(t)}{dt}=\alpha\,\Big(F(t)-R(t)\Big)\left(1-\frac{F(t)}{N}\right)
\label{eqF}
\ee
which is valid for any choice of ${\cal P}(t)$. 
\subsubsection{Standard SIR model}
\noindent If we now choose ${\cal P}(t)=e^{-\beta t}$, inserting it in Eqs.~(\ref{eqI})-(\ref{eqR}) we recover the well-celebrated SIR model \cite{anderson,keeling,sirmodel}
\bea
\label{S_sir}
&&\frac{dS(t)}{dt}=-\frac{\alpha}{N} {S(t)}I(t)\\
\label{I_sir}
&&\frac{dI(t)}{dt}= \frac{\alpha}{N}{S(t)}I(t)-\beta I(t)\\
&&\frac{dR(t)}{dt}=\beta I(t)
\label{R_sir}
\eea
where $\beta$ is the recovery rate. In order to make a comparison with what follows let us solve these equations when 
the population $N$ is very large, and as long as $F(t)\ll N$, such that $S(t)\simeq N$. In this situation we have 
\be
\frac{dI(t)}{dt}= (\alpha-\beta)\, I(t) 
\ee
and solving $\frac{dF(t)}{dt}= \alpha I(t)$, with the initial condition $F(t=0)\equiv F_o=I_o$, we get that the growth of the total number of infections, at the early stage, has the following form
\be
\label{F_sir}
F(t)=F_o\,\frac{\beta- \alpha \, e^{(\alpha-\beta)t}}{\beta-\alpha}.
\ee
\subsubsection{Delay model}
\begin{figure}[h]
  \includegraphics[width=6.3cm]{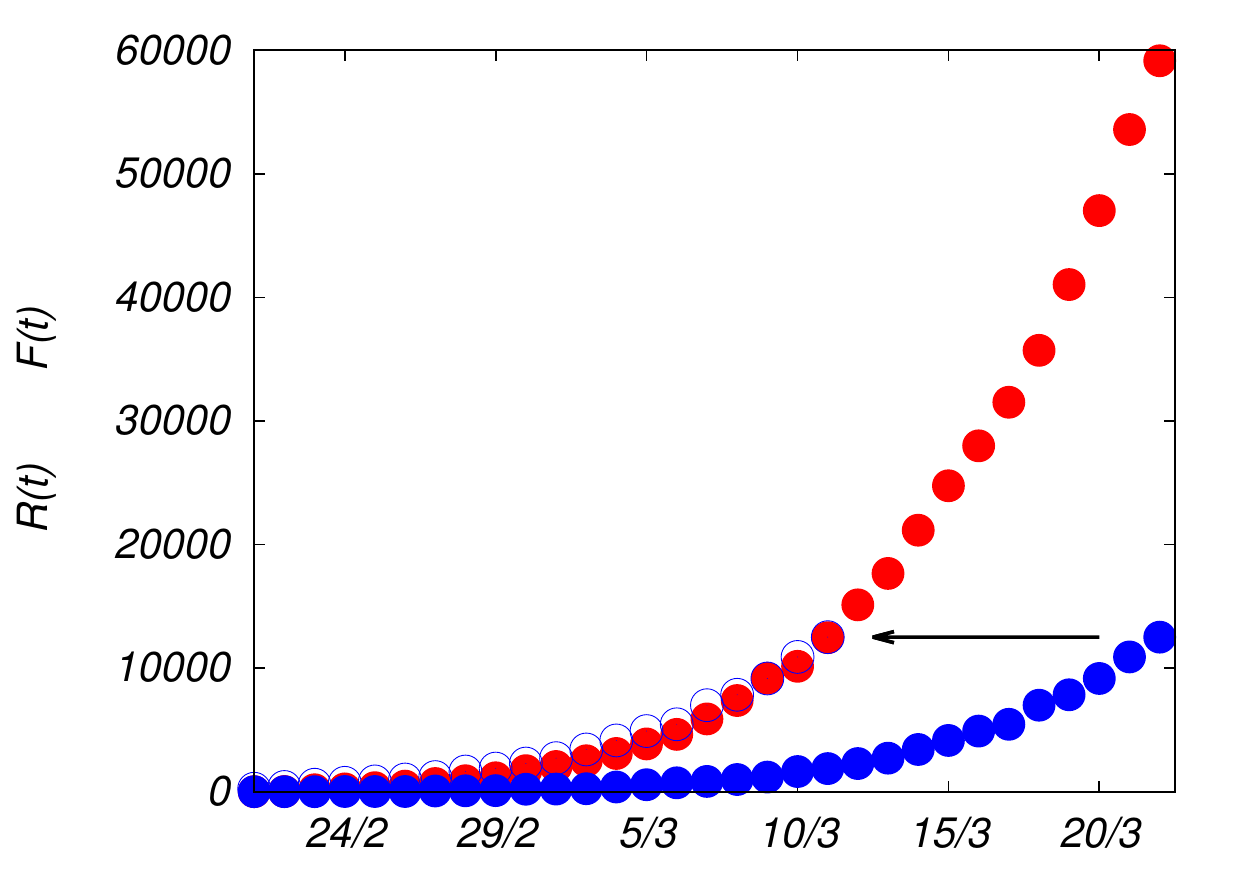}
\caption{
Total number of confirmed cases of Covid-19 in Italy, $F(t)$ (red dots), reported in Ref.~\cite{repubblica}, since $21$th February to $22$th March $2020$, compared with the closed cases, $R(t)$ (blue dots), in the same period of time. If the numbers of closed cases are shifted in time by $ T\simeq 11$ days (blue circles) they fairly overlap with the total numbers of cases.}
\label{Fconf}
\end{figure}

\noindent If, instead, we choose ${\cal P}(t)=\Theta(t-T)$, a step function, namely ${\cal P}(t)=1$ for $0\le t\le T$ and ${\cal P}(t)=0$ for $t>T$, inserting it in the Eqs.~(\ref{eqI}), (\ref{eqR}), being ${\cal P}^\prime(t-x)=-\delta(t-x-T)$, we get
\bea
\label{S_delay}
&&\frac{dS(t)}{dt}=-\frac{\alpha}{N} {S(t)}I(t)\\
\label{I_delay}
&&\frac{dI(t)}{dt}=\frac{\alpha}{N} {S(t)}I(t)-\frac{\alpha}{N} {S(t-T)}I(t-T)\\
&&\frac{dR(t)}{dt}=\frac{\alpha}{N} {S(t-T)}I(t-T)
\label{R_delay}
\eea
From the equations above it is easy to see that $\frac{dR(t)}{dt}=\frac{dF(t-T)}{dt}$, therefore $R(t)=F(t-T)+C$, with $C$ a constant value. 
We remind that, contrary to $I(t)$, either $F(t)$ and $R(t)$ are both cumulant quantities, namely they are monotonic increasing functions. Requiring that $F(t)$ saturates at $t\rightarrow \infty$, the constant value has to be $C=0$, therefore
\be
\label{delay}
R(t)=F(t-T)
\ee
This equation describes the realistic fact that the total number of cases at some time $t$ becomes that of removed cases at later time $t+T$, namely after an infectious period $T$. 
This seems to be the case also for the new coronavirus spreading, by looking at some reported data for Covid-19 in Italy, shown in Fig.~\ref{Fconf} (see also Ref.~\cite{symptoms}). Eq.~(\ref{delay}) allows us to write Eq.~(\ref{eqF}) in terms of only the function $F(t)$. 
Eq.~(\ref{eqF}), for the delay model, therefore, reads
\be
\frac{dF(t)}{dt}=\alpha\,\Big(F(t)-F(t-T)\Big)\left(1-\frac{F(t)}{N}\right)
\label{eq2}
\ee
where $F(t-T)=0$ for $t<T$. This delay differential equation is known to be linked to non-Markovian dynamics \cite{kiss}.
If we consider the case where the population $N$ is very large, and as long as $F(t)\ll N$, we can neglect the logistic term, $\big(1-\frac{F(t)}{N}\big)\simeq 1$, so to have
\be
\frac{dF(t)}{dt}=\alpha\,\Big(F(t)-F(t-T)\Big)
\label{eq_noloc}
\ee
We expect that this functional retarded differential equation, 
Eq.~(\ref{eq_noloc}), at least, at the early stage of the infection, 
could describe accurately the spreading of the epidemic disease. 
\subsubsection{Basic reproduction number}
\begin{figure}[h]
  \includegraphics[width=10cm]{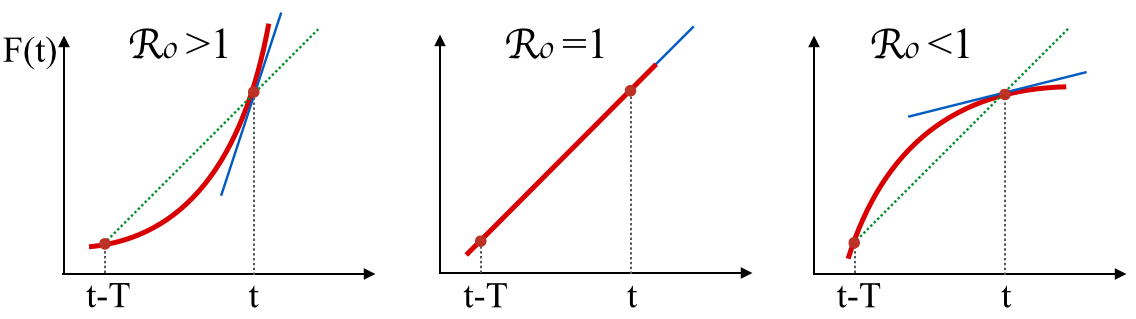}
\caption{${\cal R}_0$ for different slops of the epidemic curve as compared with its increment in a time interval $T$.}
 \label{fig_R0}
\end{figure}

\noindent
Let us rewrite Eq.~(\ref{eq_noloc}), for $t>T$, in the following form
\be
\frac{d F(t)}{d t}=  
{\cal R}_0\, \frac{F(t)-F(t-T)}{T}
\label{eq_R0F}
\ee
where we introduce and naturally identify ${\cal R}_0$ as the so-called basic reproduction number 
\be
{\cal R}_0=\alpha \,T
\ee
which is a widely used parameter for predicting whether the infectious disease will spread into a population or turns off, and represents the average number of cases originated by a single infectious case during the infectiousness period. 
Eq.~(\ref{eq_R0F}) implies that the first derivative of $F(t)$ is equal to its increment in a time interval $T$, divided by $T$, namely $F(t)$ is linear in $t$ if the rate is equal to the critical value 
$\alpha ={T}^{-1}$ (${\cal R}_0=1$).  
For $\alpha >T^{-1}$ (${\cal R}_0>1$), the function $F(t)$ increases more than linearly, while for $\alpha <T^{-1}$ (${\cal R}_0<1$), $F(t)$ goes slower than linearly (see Fig.~\ref{fig_R0}). 
If we let $\alpha$ vary in time, when $\alpha=T^{-1}$ (${\cal R}_0=1$) the function $F(t)$ has an inflection point, where it changes from being concave to convex or vice versa.
Making a comparison with the SIR model, where ${\cal R}_0=\alpha/\beta$, one can identify $\beta$, the recovery rate with the inverse of the recovery time $\beta \sim 1/T$.
Notice that ${\cal R}_0$ is well defined as long as $F(t)\ll N$, namely in the early stage of the infection. In general terms one has to define the generalized reproduction number ${\cal R}_t=\alpha (1-F(t)/N) T$ so that Eq.~(\ref{eq2}) can be written in the same form of Eq.~(\ref{eq_R0F}) with ${\cal R}_t$ instead of ${\cal R}_0$.

\subsubsection{Analytical solution}
\noindent
In this section we will provide the exact solution of Eq.~(\ref{eq_noloc}). 
Writing the time $t$ as $t=n T+t'$, where $n=\lfloor\frac{t}{T}\rfloor$ is the integer part of $t/T$, the solution of Eq.~(\ref{eq_noloc}) is
given by
\be
F(t)=F(n T+t')=F_o \prod_{\ell=1}^n A_{\ell}(T)\, A_{n+1}(t')
\label{eq_iter}
\ee
where the functions $A_\ell$ fulfill the following iterative equation
\be
\label{eq_Al}
A_\ell(t)=e^{\alpha t}\left(1- \alpha\,A_{\ell-1}(T)^{-1}\int_0^t dt' e^{-\alpha t'} A_{\ell-1}(t') \right)
\ee
with $A_0(t)=0$ for any $t<T$ and $A_0(T)=1$, so that, for $\ell=1$, we recover $A_1(t)=e^{\alpha t}$. The full exact solution is, therefore, obtained by solving a cascade of $n$ local integrals. 
The proof of Eqs.~(\ref{eq_iter}) and (\ref{eq_Al}) is given in Methods.\\
At time $t=n T$, from Eq.~(\ref{eq_Al}), performing the chain of integrals, and putting the results in Eq.~(\ref{eq_iter}), we get the following exact result 
\be
\label{Fexact}
F(n \hspace{0.01cm} T)=F_o\sum_{\ell=0}^{n}\frac{(-1)^\ell}{\ell!}\big((n-\ell)\,\alpha\hspace{0.01cm}T\big)^\ell e^{(n-\ell) \alpha T} 
\ee
For instance, for $n=1$ and $n=2$, namely up to twice the infectiousness period, the total number of cases is simply 
$F(n\hspace{0.01cm}T)=F_o\left(e^{n \alpha T}-(n-1)\,\alpha T \,e^{(n-1)\alpha T}\right)$.  
Surprisingly we find that Eq.~(\ref{Fexact}) depends only on $(\alpha \hspace{0.02cm} T)$, which is the basic reproduction number ${\cal R}_0$.  It is easy to check from Eq.~(\ref{Fexact}) that, while for large ${\cal R}_0=\alpha T$, $F(nT)$ is dominated by an exponential behavior, for ${\cal R}_0=1$, $F(nT)$ becomes linear in $n$. 
From Eqs.~(\ref{eq_iter}) and (\ref{eq_Al}) we can also write the following equation
\be
F(t)=F(n \hspace{0.03cm} T+t')=e^{\alpha t'}\left(F(n \hspace{0.03cm} T)-
\alpha \int_0^{t'} d s \, e^{-\alpha s} F\big((n-1) \hspace{0.03cm} T+s\big)
\right)
\label{Fex_generic}
\ee
By iteration one gets simply
\be
F(n \hspace{0.03cm} T+t')=e^{\alpha t'}\sum_{m=0}^{n} (-\alpha)^m {\cal I}_{m}(t') \,F\big((n-m)\hspace{0.03cm}T\big) 
\ee
where ${\cal I}_m$ fulfills the following recursive equation, with inital value ${\cal I}_0=1$,  
\be
{\cal I}_m(t')=\int_0^{t'}ds\, {\cal I}_{m-1}(s)=\frac{{t'}^m}{m!}
\ee
The final exact result for any time is, therefore,
\bea
\nonumber 
F(t)=F(n \hspace{0.03cm} T+t')&=&e^{\alpha t'} \sum_{m=0}^{n}\frac{(-1)^m}{m!}(\alpha t')^mF\big((n-m)\hspace{0.03cm}T\big)\\
&=&F_o\,e^{\alpha t'}  \sum_{m=0}^{n} \sum_{\ell=0}^{n-m}\frac{(-1)^{\ell+m}}{\ell!\,m!}(\alpha t')^m\big((n-m-\ell)\,{\cal R}_0\big)^\ell  e^{(n-m-\ell) {\cal R}_0} 
\label{final_result}
\eea
where $t'=\textrm{mod}(t,T)$. For practical reasons, in order to avoid indeterminate forms, for $t'=0$ and $m=0$, in Eq.~(\ref{final_result}) one can add an infinitesimal term $\epsilon\rightarrow 0$, so to have $(\alpha t'+\epsilon)^m$.  
Once we have the total number of infections $F(t)$ at any time, we get also the number of removed case, $R(t)=F(t-T)$, and  
we can easily calculate, from Eq.~(\ref{final_result}), the number of persons who are still infected, at a given time $t$, which, by definition and from Eq.~(\ref{eq_noloc}), is given by
\be
\label{eq_It}
I(t)=F(t)-R(t)=F(t)-F(t-T)=\frac{1}{\alpha}\frac{dF(t)}{dt}.
\ee

\subsubsection{Comparison between the delay model and the standard SIR model}

As we have seen, one assumption the standard SIR model is based on is that the time in which individuals remain infectious is described by an exponential distribution, which is however biologically rather unrealistic. 
In reality, infectious periods are fairly closely centered about the mean duration of an infection. A constant infectious period is therefore a more realistic assumption.
The conventional SIR model being formulated in terms of 
ordinary differential equations, requires the presence of an effective recovery (and fatality) rate
which might not correspond to the actual rate since the new cases of recovery (and fatality) come from infected cases occurring a few days earlier. 
For that reason, instead of writing the problem in terms of ordinary differential equations one has to do it in terms of 
functional differential equations, as for the delay model. 
Even if the recovery rate of the SIR model is chosen to be equal to the inverse of the average infectious period, the dynamics obtained by solving Eqs.~(\ref{S_sir}-\ref{R_sir}) does not correspond to the dynamics obtained by solving Eqs.~(\ref{S_delay}-\ref{R_delay}). 
\begin{figure}[h!]
\subfigure[]
 {\includegraphics[width=6.cm]{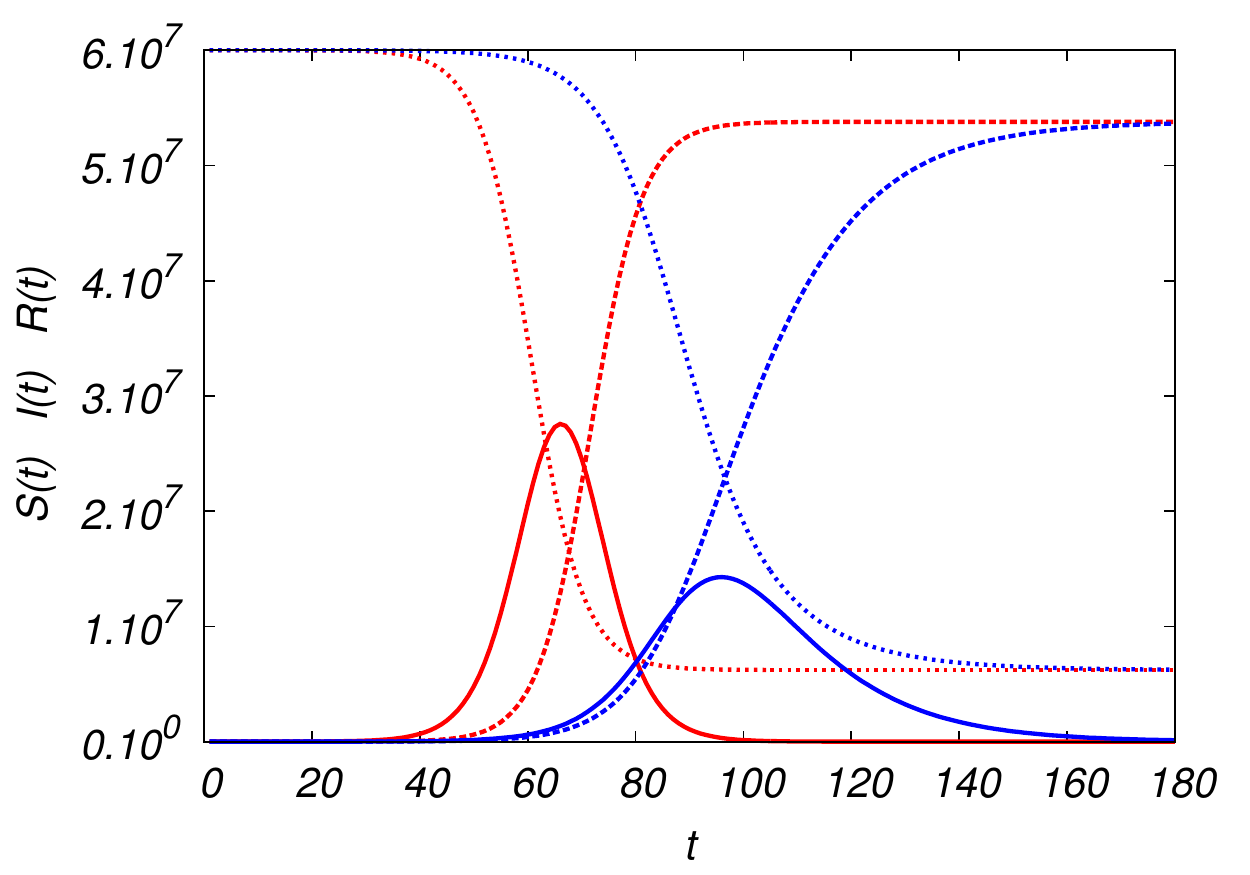}}\hspace{0.3cm}
 \subfigure[]
{\includegraphics[width=6.cm]{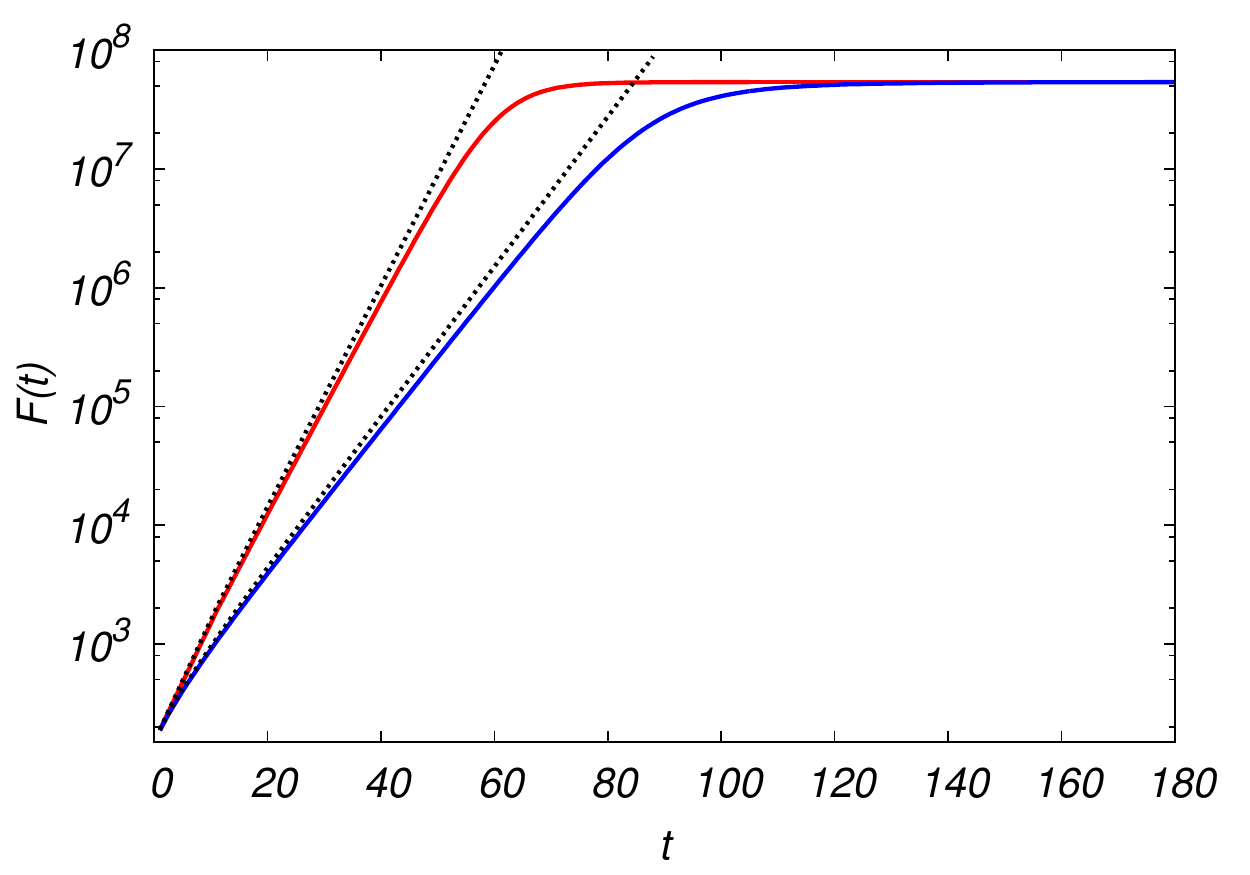}}
\caption{(a) Number of susceptible $S(t)$ (dotted lines), infected $I(t)$ (solid lines), and recovered $R(t)$ (dashed lines) persons as functions of time, for the SIR model, Eqs.~(\ref{S_sir}-\ref{R_sir}) (blue lines) and for the delay model, Eqs.~(\ref{S_delay}-\ref{R_delay}) (red lines), with initial conditions $I_o=I(t=0)=150$ and $R(0)=0$, and $S(t)+I(t)+R(t)=N=6\times 10^7$, with parameters $\alpha=0.23$ per unit of time and $T=\beta^{-1}=11$ units of time (e.g. days), therefore the basic reproduction number in both the models is ${\cal R}_0\simeq 2.5$. (b) Total number of infected persons $F(t)=I(t)+R(t)$, in log-scale, as a function of time from the standard SIR model (blue solid line) and from the delay model (red solid line). The gray dotted lines are the analytical results from Eq.~(\ref{F_sir}) for the SIR model and Eq.~(\ref{final_result}) for the delay model, valid in the first stage of the infection.}
\label{comp}
\end{figure}
As shown in Fig.~\ref{comp}, even with the same initial conditions and the same ${\cal R}_0$, the growth and the expected peak of the spreading of the infectious disease are quite different between the two models, even if the asymptotic final values are the same. For ${\cal R}_0\simeq 2.5$ the SIR model predicts a much lower peak of $I(t)$ with respect to that expected from the delay model, which is much sharper and occurs much earlier. In other words, the outbreak of an epidemic disease might be underestimated by the standard SIR model. We notice also that the analytic expression for $F(t)$ in Eq.~(\ref{final_result}) describes fairly well the increase of the infection, at least in its early stage.

\subsubsection{Time-dependent infection rate: analytical solution}
Let us now consider the possibility of having a time-dependent infection rate $\alpha(t)$ in the dynamical equation for the total number of infected persons
\be
\frac{dF(t)}{dt}=\alpha(t)\,\Big(F(t)-F(t-T)
\Big).
\label{eq_Frt0}
\ee
Also in this more general case the exact solution, valid for any profile of $\alpha(t)$, can be written in the same form of Eq.~(\ref{eq_iter}), namely, $F(t)=F(n \hspace{0.02cm}T+t')=F_o \prod_{\ell=1}^n A_{\ell}(T)\, A_{n+1}(t')$, 
where now the functions $A_\ell$ are given by
\be
\label{eq_Al_rt}
A_{\ell+1}(t)=e^{\int_{\ell  T}^{\ell T+t} \alpha(t') dt'}\left(1- A_{\ell}( T)^{-1}\int_0^t dt' \,\alpha\big(\ell \,T+t'\big)e^{-\int_{\ell T}^{\ell T+t'} \alpha(t'') dt''} A_{\ell}(t') \right).
\ee
For instance, $A_1(t)=e^{\int_{0}^{t} \alpha(t') dt'}$, $A_2(t)=e^{\int_{ T}^{ T+t} \alpha(t') dt'}\left(1- e^{-\int_{0}^{ T} \alpha(t') dt'}
\int_0^t dt' \,\alpha\big( T+t'\big)e^{-\int_{ T}^{ T+t'} \alpha(t'') dt''}e^{\int_{0}^{t'} \alpha(t'') dt''}   \right)$ and so on. For constant $\alpha$, Eq.~(\ref{eq_Al_rt}) reduces to Eq.~(\ref{eq_Al}). 
See Methods for more details about the derivation. 
The solution $F(t)$ has therefore to fulfill the following recursive equation, after splitting the time in $n$ intervals $ T$ with the residual time $t'=\textrm{mod}(t,T)$
\be
F(t)=F(n \hspace{0.03cm} T+t')=e^{\int_{n T}^{n T+t'} \alpha(s) ds}\left(F(n \hspace{0.03cm} T)-
\int_0^{t'} d s \, \alpha(n T+s)e^{-\int_{n T}^{n T+s} \alpha(t'') dt''} F\big((n-1) \hspace{0.03cm} T+s\big)
\right)
\label{Fex_generic}
\ee
This general result implies that if we knew the time dependence of the infection rate or if we could tailor its evolution by, for instance, containment measures, we can know the exact analytical expression of $F(t)$, the total number of infected persons, as a function of time, as long as $F(t)$ is much smaller than $N$.

\subsection{Covid-19 in Italy}

\begin{figure}[h]
  \includegraphics[width=6.3cm]{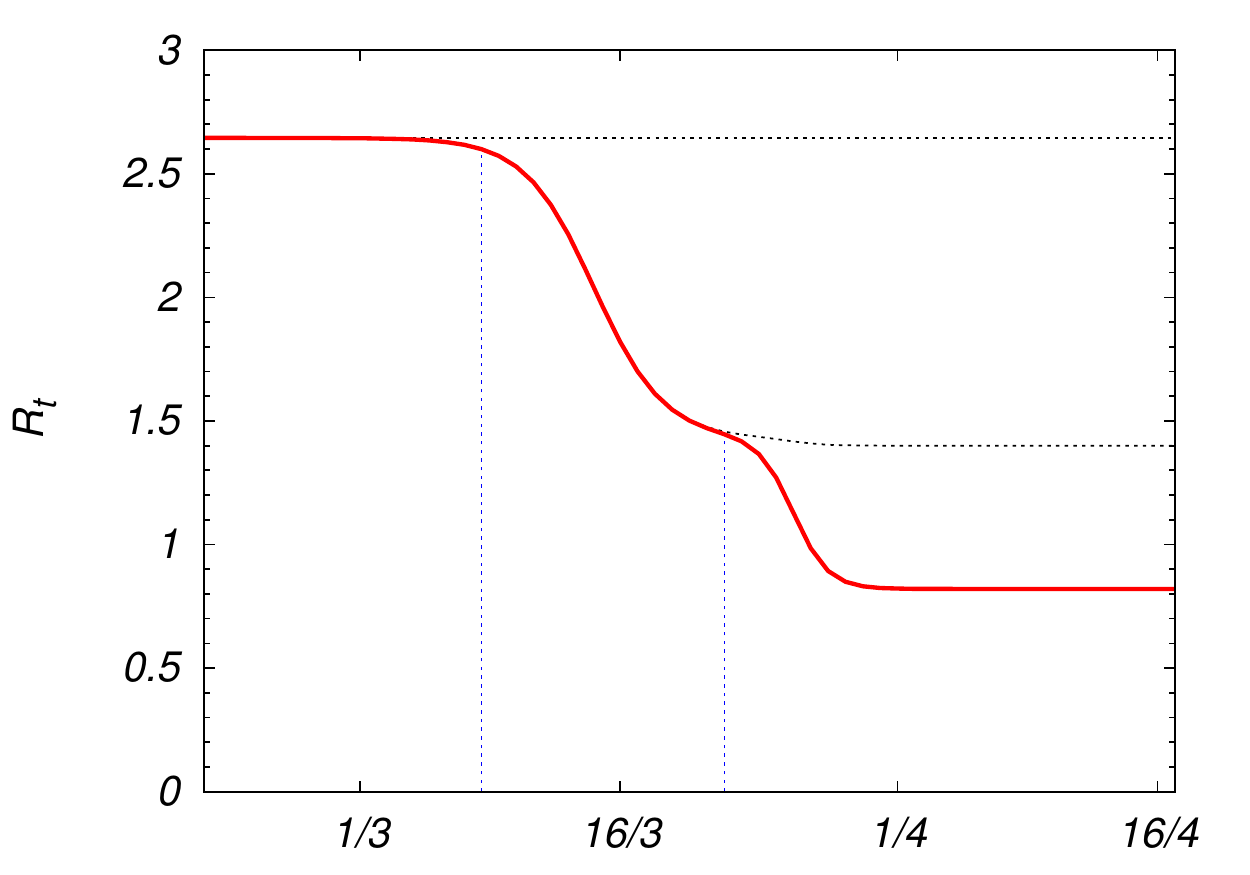}
\caption{Time-dependent reproduction number ${\cal R}_t= \alpha(t) T$, as a function of time, based on the profile for the infection rate described by Eq.~(\ref{eq_rt}). We take $ T$ about $11$ or $12$ days \cite{symptoms},  
$t_1$ between  $13$th and $14$th March $2020$,  $t_2$ on $26$th March, $\tau_1\sim 2$ days, $\tau_2\sim 1$ day. 
The initial value is ${\cal R}_0=\alpha_1 T \simeq 2.65$ (in agreement with other estimates, see e.g. \cite{gaeta}), the intermediate value is ${\cal R}_t=\alpha_2 T \simeq 1.45$, and the final value, ${\cal R}_t=\alpha_3 T\simeq 0.85$. The vertical dotted blue lines point the dates of the main laws for the containment measures ($8$th-$11$th March and $22$th March 2020).
The gray dotted lines correspond to ${\cal R}_t$ in the absence of the first and the second containment measures.}
\label{rt}
\end{figure}

Let us consider the delay model in its general form, Eq.~(\ref{eq2}) where the infection rate $\alpha$ varies in time 
\be
\label{eq_Frt}
\frac{dF(t)}{dt}= \alpha(t)\,\Big(F(t)-F(t- T)
\Big)\left(1-\frac{F(t)}{N}\right)
\ee
as the effect of some containment measures taken in order to reduce the impact of an infection on the population.
 As an example, let us suppose that 
$ \alpha(t)$ is modified by social distancing measures, lockdown and the shutdown of many work activities, as it is happening in Italy (and in many other
countries) to mitigate and reduce the spreading of the new coronavirus, Covid-19, after two main decrees imposed by the Italian Prime Minister ordering the lockdown of the whole national territory, taken on March $11$-th (lockdown and shutdown of many stores) and March $22$-th 2020 (shutdown of many factories and strengthening of social distancing measures), and after some other measures taken right before for local regions (e.g. the decree of March $8$-th for the lockdown of Lombardy and other areas). As a result, we can imagine that $ \alpha(t)$  decreases smoothly after those dates taking into account the adaptation time for the individuals to the new social behaviors and the period needed to complete the last activities before the blockade of the factories. Let us suppose, therefore, that $ \alpha(t)$ can change in time according to a smooth step function as in Eq.~(\ref{eq_rt}), 
\be
\label{eq_rt}
 \alpha(t)=\left(\frac{\alpha_1-\alpha_2}{1+e^{(t-t_1)/\tau_1}}+\alpha_2-\alpha_3\right)\frac{1}{1+e^{(t-t_2)/\tau_2}}+\alpha_3
\ee
where $t_1$ and $t_2$ are the times where the steps are located, $\tau_1$ and $\tau_2$ make  the function to be smooth, $\alpha_1$ is  
the initial observed infection rate which causes the starting exponential growth of the epidemic disease, $\alpha_2$ the intermediate rate, which fits with the data, supposed to be reached after the first decree of lockdown, and $\alpha_3$ the supposed asymptotic infection rate after the second decree of lockdown. 
Fixing the average of recovery and fatality time $ T$, the reproduction number is also a function of time, therefore we define 
\be{\cal R}_t= \alpha(t) T
\ee 
with a  profile shown in Fig.~\ref{rt}.  More precisely ${\cal R}_t= \alpha(t)(1-F(t)/N)T$, but as we will see, because of the containment measures, $F(t)\ll N$ at any time.

\begin{figure}[h!]
\subfigure[]
{  \includegraphics[width=6.3cm]{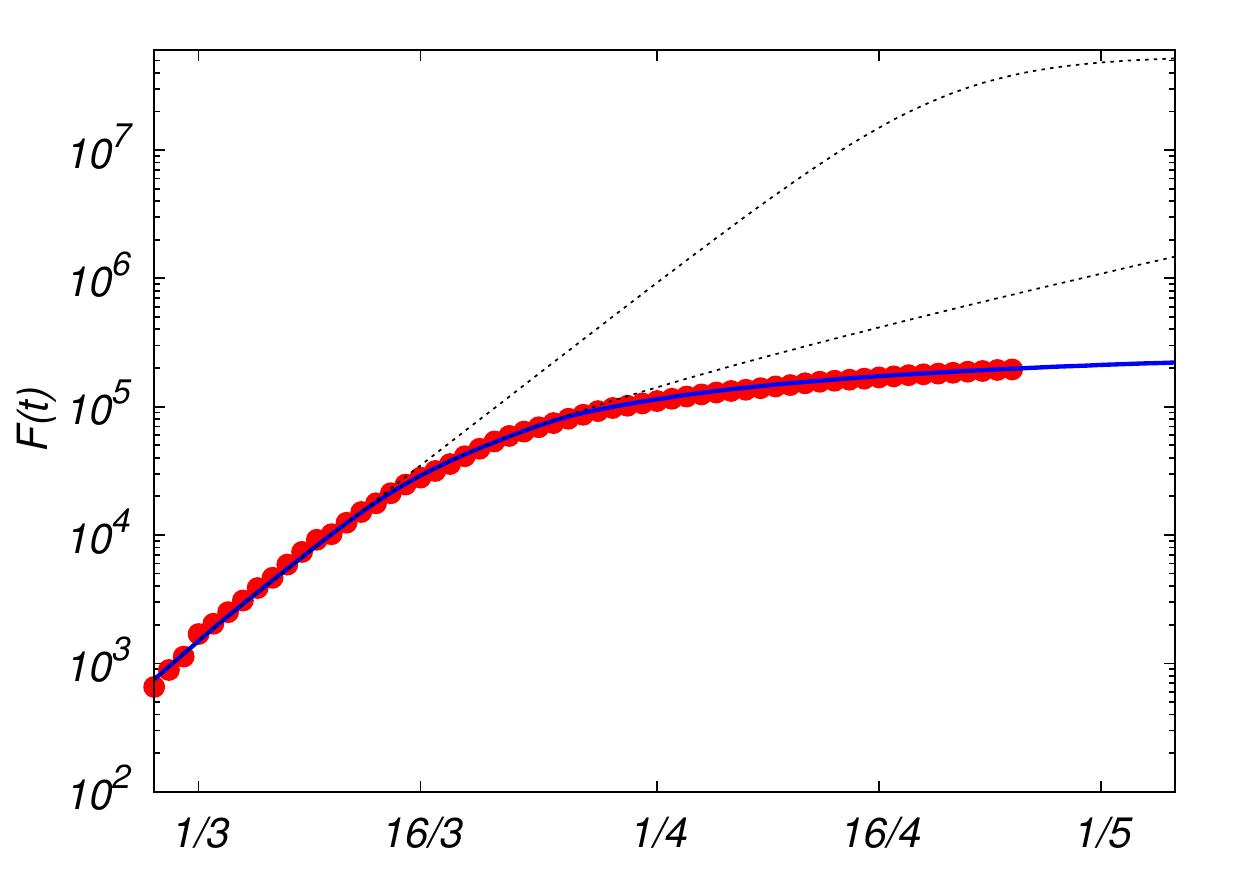}}\hspace{0.3cm}
\subfigure[]
{    \includegraphics[width=6.3cm]{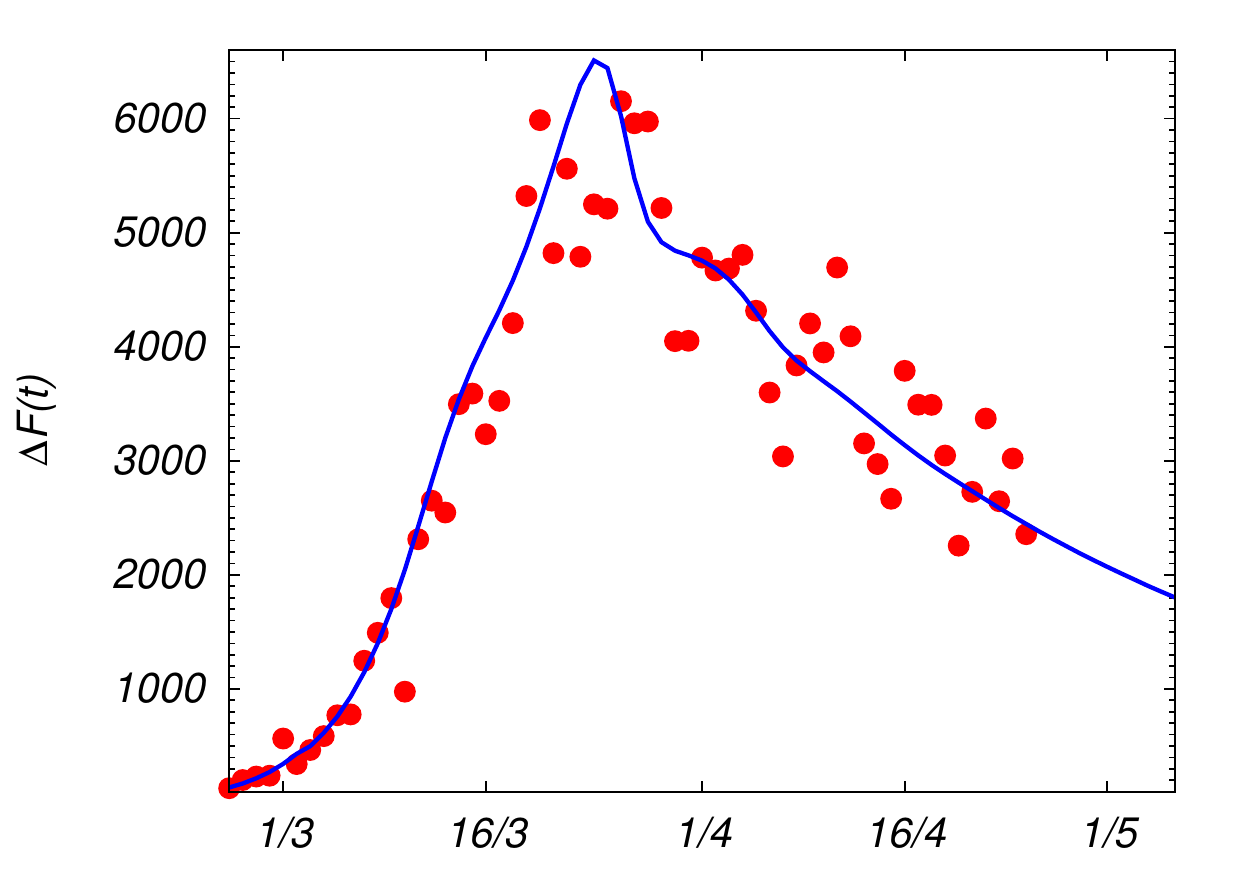}}
\caption{(a) Total number of infected persons over time, $F(t)$ (red points), from official data for Covid-19 in Italy \cite{repubblica}, where $N=6\cdot10^{7}$, up to $25$th April. 
The blue line is the theoretical prediction $F(t)$ as solution of Eq.~(\ref{eq_Frt}), with initial conditions, fixed at $t=0$ the $21$th February, $F_o\simeq 150$, and $\alpha=\alpha_1\simeq 0.23$ (${\cal R}_0\simeq 2.65$), and using the profile for the infection rate given by Eq.~(\ref{eq_rt}), with the parameters reported  in Fig.~\ref{rt}. The gray dotted lines are the expected curves for $F(t)$ if the first and the second containment measures ($8$-$11$th March and $22$th March) had not been taken. (b) Daily number of infected persons, $\Delta F(t)$, compared with the theoretical result obtained performing $\frac{dF(t)}{dt}$ from the solution of Eq.~(\ref{eq_Frt}).}
\label{fig_theory}
\end{figure}

\begin{figure}[h]
\subfigure[]
{   \includegraphics[width=6.3cm]{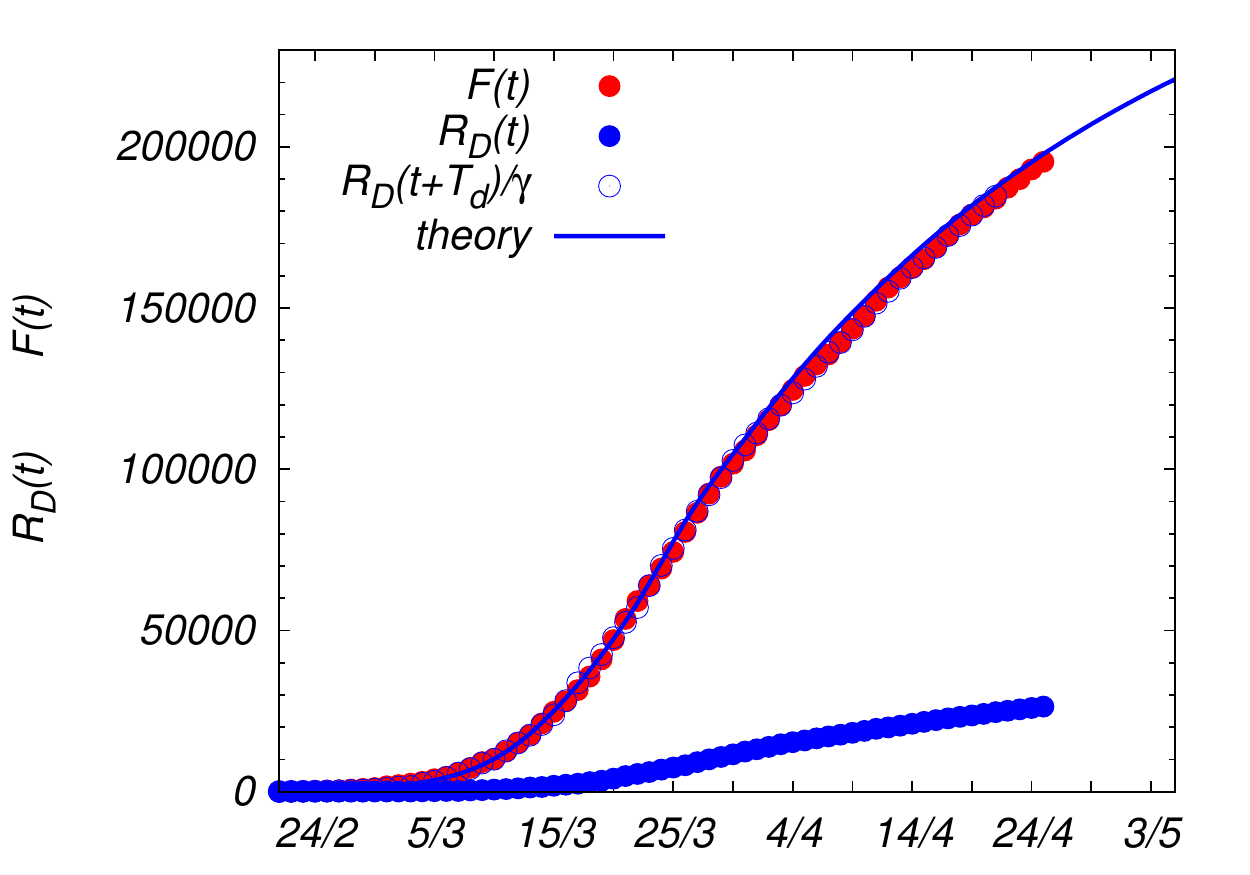}}\hspace{0.3cm}
\subfigure[]
{      \includegraphics[width=6.3cm]{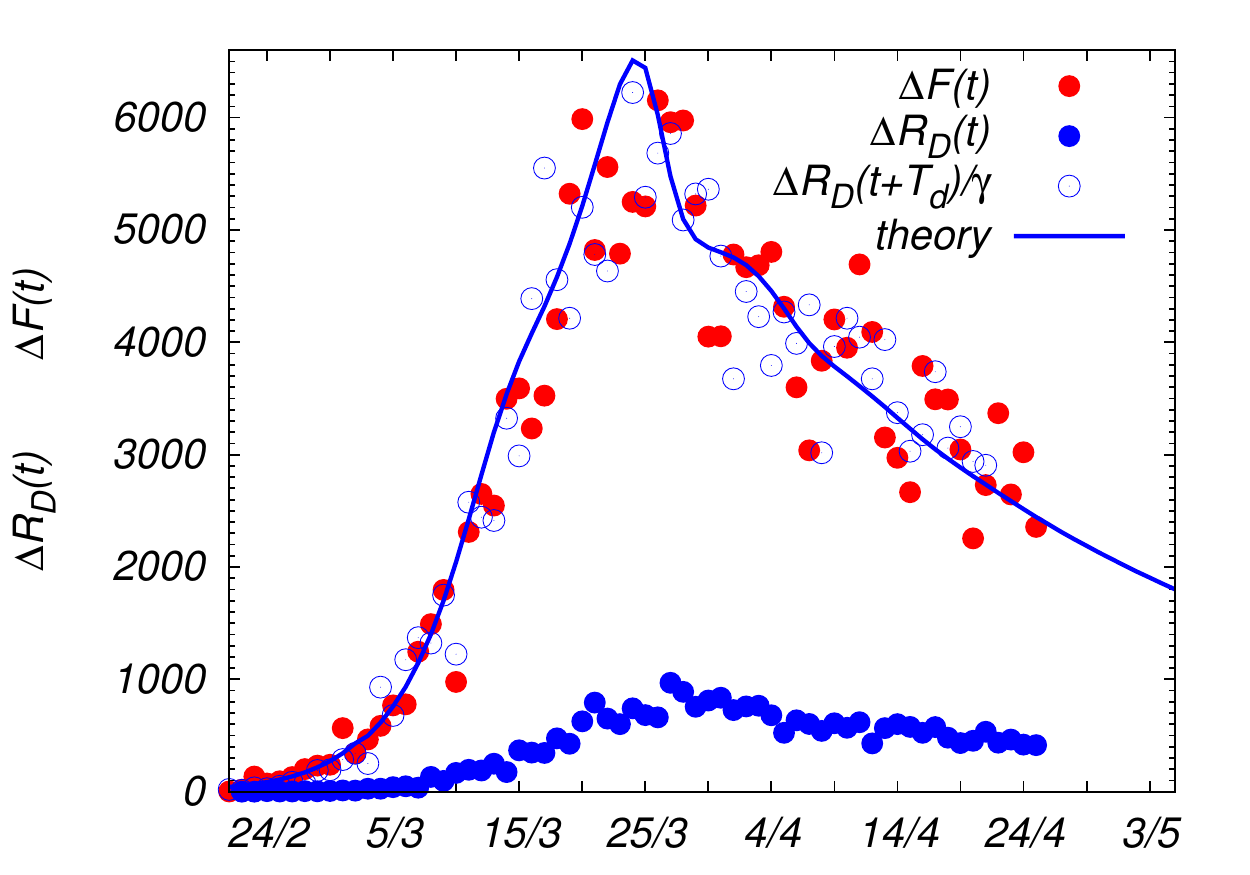}}
\caption{
(a) Total number of confirmed cases of Covid-19 in Italy, $F(t)$ (red dots), reported in Ref.~\cite{repubblica}, up to $25$th April $2020$, compared with the deceased cases, $R_D(t)$ (blue dots), in the same period of time. If the numbers of dead persons are shifted in time by $ T_d\simeq 4$ days and rescaled by $\gamma^{-1}=7$ (blue circles) they perfectly overlap with the total numbers of cases. The blue solid line is theoretical prediction for $F(t)$, as solution of Eq.~(\ref{eq_Frt}). (b) Daily number of infected persons, $\Delta F(t)$ (red dots), and daily number of victims $\Delta R_D(t)$ (blue dots). 
The blue circles are the daily number of victims after rescaling according to Eq.~(\ref{rescale}), with $ T_d=4$ days and $\gamma=1/7$.
The solid blue line is $\frac{dF(t)}{dt}$ from the solution of Eq.~(\ref{eq_Frt}).}
\label{FRd}
\end{figure}

\noindent
Solving Eq.~(\ref{eq_Frt}), or, analogously, using the recursive relation in Eq.~(\ref{Fex_generic}), with the time-dependent rate $ \alpha(t)$ 
given by Eq.~(\ref{eq_rt}), with the parameters reported in Fig.~\ref{rt}, 
we obtain the solution $F(t)$ which slowly goes to saturation over time, in perfect agreement with the data for the total number of confirmed infected cases, as shown by Fig. \ref{fig_theory}, where the blue line is the expected curve, while the red points are the official data. 
The dotted gray lines in Fig.~\ref{fig_theory} represent $F(t)$ if the containment measures had not been taken. 
As one can see from Figs.~\ref{rt}-\ref{fig_theory}, only when ${\cal R}_t$ becomes smaller than $1$, the curve flattens allowing for a stop of the epidemic spreading,  avoiding that a large part of the population gets infected. For ${\cal R}_t\simeq 1$, $F(t)$ would increase linearly, and $I(t)$ would become almost constant, meaning that the number of new infections would be equal to the number of closed cases. 
A reliable forecast has to take into account the fact that the official data of infectious cases are made by counting 
mostly the symptomatic cases, probably discarding other infectious cases which could transfer the virus even without or with mild symptoms. Moreover, the data of both the total number of infected persons and that of the recovered ones could be affected by the procedure, the realization times and the number of the diagnostic tests. However, since our model relies on the infectiousness time, it does not need a fitting of the data for recovered persons which may be affected by systematic errors. This uncertainty on the data for closed cases would compromise the result for the SIR model. On the contrary, our theoretical prediction based on the delay model agrees fairly well with the data-set for total infected cases, as shown in Fig. \ref{fig_theory}. \\
As a final remark we remind that most of the confirmed infected cases in Italy are counted after the appearance of the symptoms and the persons who exhibit severe ones are mostly hospitalized, and afterwards counted as infected persons. Some of them, unfortunately, die approximately $4$ days after (therefore after approximately $9$ days from the onset of the first symptoms, as reported by the Istituto Superiore di Sanit\`a \cite{ms}). 
We observe that, splitting the closed cases between real recovered persons, $R_R$, and dead persons, $R_D$,
\be
R(t)=R_R(t)+R_D(t)
\ee
and since the confirmation of recovery needs extra diagnostic tests which are not widely performed yet, the most reliable data are those related to dead persons $R_D(t)$, which are found to be linked to the total number of confirmed infected cases, $F(t)$, in the following way
\be
\label{rescale}
R_D(t)\simeq \gamma\, F(t- T_d)
\ee 
with $\gamma=\frac{1}{7}$ and a delay time of $ T_d=4$ days, as show in Fig.~\ref{FRd}. 
The number of victims follows the number of total confirmed cases and it is equal to $1/7$ of its value four days before. \\
The fatality of the sick persons, those who exhibit some symptoms, is therefore quite high, $\lim_{t\rightarrow\infty}\frac{R_D(t)}{F(t)}=\gamma\simeq 14\%$. 

\section{Discussion}
We present a simple but realistic model for describing epidemic spreading, based on the fact that the closed cases come from infected ones at an early time. This observation allows us to formulate the problem in terms of a single functional differential equation depending on two well defined clinically relevant parameters: the infection rate and the infectiousness time. We provide the exact analytical solution for such an equation, in the limit of a large population, finding how it depends  on the basic reproduction number ${\cal R}_0=\alpha T$, see Eqs.~(\ref{Fexact}) and (\ref{final_result}). Contrary to the result of the conventional SIR model, the total number of cases has a combined polynomial and exponential growth.  
We derive the analytic solution also in the presence of a generic time-dependent infection rate, which is the case when some measures are taken to weaken the spreading of the epidemic disease. 
We apply, therefore, our model to study the spreading of Covid-19 in Italy, allowing the infection rate to vary in time, as a result of some containment measures implemented by the government in order to mitigate the consequences of the infection on the population. 
We find perfect agreement between the official data and the expected theoretical results. 
In general terms, the reproduction number should be suppressed well below $1$ in order to rapidly recover the initial condition. By a rough estimation, in order to have a decline of the infection as fast as its growth, containment measures or possible therapies should be so effective to reduce the basic reproduction number and reach the final value ${\cal R}_{f}$ such that 
${\cal R}_{f}\simeq \frac{{\cal R}_{0}}{2{\cal R}_{0}-1}$, 
starting from an initial value ${\cal R}_{0}$. 
In the case of Covid-19 in Italy, the initial value for the basic reproduction number was ${\cal R}_{0}\simeq 2.6$, while the current one (April) seems to settle at ${\cal R}_{f}\simeq 0.8$, implying a rather slow decline of the infection. 
Finally we discussed the fatality rate, showing that the number of victims is exactly a fraction of the total number of cases few days before.  
Before we conclude a final comment is in order. The confirmed cases are mostly symptomatic or mild symptomatic. There are also asymptomatic cases which may contribute to the spreading of the infection. However, by scaling arguments, the infection rates of the symptomatic and asymptomatic are expected to be equal, otherwise either symptomatic or asymptomatic cases might become irrelevant. Under the hypothesis that the infectiousness time does not depend on the strength of the symptoms, the ratio between the total number of asymptomatic and symptomatic cases should be constant, although it could be very large. As a result, the total number of infected persons should be equal to the number of symptomatic cases times an overall pre-factor greater than one. The conclusion is, therefore, that, as far as the time evolution of the infection is concerned, which is the aim of this work, the study of only symptomatic cases is still relevant and greatly meaningful.

\section{Methods}%
\subsection{Solution of the retarded differential equation}
\label{ApA}
For $t\le T$, the solution of Eq.~(\ref{eq_noloc}) is $F(t)=F_o e^{\alpha t}$. Let us consider $t=T+dt$ with infinitesimal $dt$, from Eq.~(\ref{eq_noloc})
\be
F(T+dt)= F(T) + dt\, \alpha \left(F(T)+F(0)\right)= F(T) \left(1+\alpha  \,dt\right)-F_o \alpha  \,dt
=F_o e^{\alpha  T}(1+\alpha  \,dt)-F_o \alpha  \,dt
\ee
Using this result we can calculate
\be
F(T+2 dt)= F(T+dt) + dt\, \alpha \left(F(T+dt)+F(dt)\right)
= F_o e^{\alpha  T}(1+\alpha  \,dt)^2-F_o \alpha  \,dt\left[(1+\alpha  \,dt)+e^{\alpha  dt}\right]
\ee
Analogously, from that, we can proceed calculating
\be
F(T+3 dt)= F(T+2 dt) + dt\, \alpha \left(F(T+2 dt)+F(2dt)\right)
= F_o e^{\alpha  T}(1+\alpha  \,dt)^3-F_o \alpha  \,dt\left[(1+\alpha \,dt)^2+e^{\alpha  dt}(1+\alpha \,dt)+e^{2 \alpha  dt}\right]
\ee
and going on by adding infinitesimal time steps, we find iteratively that
\bea
F(T+ m \,dt)&=&F_o e^{\alpha  T}(1+\alpha  \,dt)^m-F_o \alpha  \,dt \sum_{j=0}^{m-1} e^{j \alpha  dt}(1+\alpha  \,dt)^{m-j-1}
\equiv F_o A_1(T) A_2(m\,dt) \\
&=& F(T) \,A_2(m\,dt)
\eea
with $A_1(T)=e^{\alpha  T}$ and defining 
\be
A_2(m\,dt) =(1+\alpha  \,dt)^m-e^{-\alpha T}\alpha  \,dt \sum_{j=0}^{m-1} e^{j \alpha  dt}(1+\alpha \,dt)^{m-j-1}.
\ee 
In particular, for $m\, dt=T$, we have an expression for $F(2T)$ in terms of the function at early time, 
$F(2T)=F(T)A_2(T)$. 
We can now start again with the iteration
\be
F(2T+dt)= F(2T) (1+\alpha  \,dt)-\alpha  \,dt F(T)=F(T)A_2(T)\left(1+\alpha  \,dt\right)-\alpha  \,dt F(T)
\ee
One can proceed in the same way as before getting 
\be
F(2T+m\,dt)=F(T)\left[(1+\alpha  \,dt)^m A_2(T) - \alpha  \,dt \sum_{j=0}^{m-1} A_2(j\,dt)(1+\alpha  \,dt)^{m-j-1}\right]
\ee
which can be written as 
\be
F(2T+m\,dt)=F(T)\,A_2(T) \,A_3(m\,dt)=F(2T) \,A_3(m\,dt)
\ee
where
\be
A_3(m\,dt)=(1+\alpha  \,dt)^m-A_2(T)^{-1} \alpha  \,dt \sum_{j=0}^{m-1} A_2(j \,dt)(1+\alpha  \,dt)^{m-j-1}.
\ee
We can notice that at any step $T$ we can perform the same calculation since we can factorize the function $F$ as
\be
F(n\,T+ m\,dt)=F(n\,T) \, A_{n+1}(m\,dt)
\ee
where, therefore, $F(n\,T)=F_o\prod_{\ell=1}^{n}A_{\ell}(T)$ and 
\be
\label{Al_discrete}
A_{\ell}(m\,dt)=(1+\alpha  \,dt)^m\left[1-A_{\ell-1}(T)^{-1} \alpha \,dt \sum_{j=0}^{m-1} \frac{A_{\ell-1}(j \,dt)}{(1+\alpha  \,dt)^{j+1}}\right].
\ee
In the continuum limit, $d t\rightarrow 0$ and $m\rightarrow \infty$, keeping finite the time interval $m \,dt=t$, reminding that
\be
\lim_{m\rightarrow \infty}\left(1+\frac{\alpha  t}{m}\right)^m =e^{\alpha t}
\ee
we finally obtain the result reported Eq.~(\ref{eq_Al}).\\
In the presence of time dependent infection rate, splitting again the time in $n$ intervals $T$ and the residual time in $m$ infinitesimal intervals $dt$, we define 
\be
\alpha (t)=\alpha (n\,T+m \,dt)\equiv \alpha ^{(n)}_{m}. 
\ee
Proceeding iteratively as done for the constant rate case, but now taking trace of the different values of $\alpha $, 
\be
F\big(n\,T +m\,dt\big)=F\big(n\,T +(m-1)dt\big)\left(1+\alpha ^{(n)}_m dt\right)-\alpha ^{(n)}_m dt \,F\big((n-1)\,T +(m-1)dt\big)
\ee
after several steps, similar to those done previously, we find that Eq.~(\ref{Al_discrete}) can be generalized in the following way
\be
A_{\ell}(m\,dt)=\prod_{i=1}^m\left(1+\alpha ^{(\ell-1)}_i \,dt\right)-A_{\ell-1}(T)^{-1} dt \sum_{j=0}^{m-1} 
\left[{\alpha ^{(\ell-1)}_{j}A_{\ell-1}(j \,dt)}\prod_{i=1}^{j}{\left(1+\alpha _i^{(\ell-1)}\,dt\right)^{m-j-1}}\right],
\ee
whose continuum limit is given in Eq.~(\ref{eq_Al_rt}).

\end{document}